\documentclass[12pt]{article}

\begin{document}

\begin{center}

\noindent{\large\bf ac hopping conduction at extreme disorder takes place on 
the bond invasion percolation cluster}

\bigskip
Muhammad Sahimi\\
{\it Mork Family Department of Chemical Engineering \& Materials Science, 
University of Southern California, Los Angeles, California 90089-1211}

\end{center}

\bigskip
It has been suggested that ac conduction in extremely disordered solids occurs
on the critical percolation cluster. In this note, we argue that in fact the
transport process takes place on the bond invasion percolation cluster (BIPC).
The structure of the BIPC is universal independent of the conductances of the 
bonds, which explains why the rescaled ac conductivity is a universal function
of the rescaled frequency. It also explains why the effective-medium 
approximation provides quantitatively accurate predictions for the effective 
conductivity.

\bigskip

\bigskip

Schr$\phi$der and Dyre (SD) [1] studied the problem of ac conduction in 
extremely disordered solids, and addressed two important issues: (1) 
experimental data [2] and numerical simulations [1,3,4] indicate that the 
rescaled ac conductivity, $\tilde{\sigma}=\sigma(\omega)/\sigma(0)$, may be a 
{\it universal} function of the rescaled frequency, $\tilde{\omega}=a\omega/
\sigma(0)$. Here, $a$ is a constant. (2) The universal ac conductivity curve 
is expressed by,
\begin{equation}
\ln\tilde{\sigma}=(i\tilde{\omega}/\tilde{\sigma})^{d_f/2}\;, 
\end{equation}
where $d_f$ is a fractal dimension associated with the cluster in which ac 
conduction occurs. In the effective-medium approximation (EMA) [5-7], one
has [8], $d_f=2$. SD assumed that ac conduction occurs in the {\it critical 
percolation cluster} (CPC) at the percolation threshold $p_c$, and suggested, 
$d_f=d_H$, where $d_H$ is the harmonic (fracton) dimension [9] that has the 
same approximate value of 4/3 in both two- and three-dimensional (3D) CPC. 
Numerical evidence presented by SD indicated that Eq. (1) with $d_f=d_H=4/3$ 
provides accurate representation of the ac conductivity of a simple-cubic 
lattice.

In this note, we (1) argue that the conduction cluster is {\it not} the CPC 
and, thus, $d_f$ cannot be equal to its $d_H$; (2) explain why $\tilde{\sigma}$
is universal, and (3) explain why the EMA provides a good approximation to 
$\tilde{\sigma}$.

SD studied ac conduction by the random barrier (RB) model. It is well-known
that in a system in which the hopping rates vary over many orders of magnitude,
conduction occurs only in a very small portion of the system. In fact, in the 
RB model ac conduction occurs in the {\it minimum spanning tree} (MST). Let 
$e_{ij}$ be the energy, or barrier, associated with bond $ij$ of a lattice. The
MST is a cluster that visits every site in the lattice such that the total
energy, $E=\sum_{ij}e_{ij}$, is {\it minimum}, with the constraint that visit 
to any site cannot create a closed loop. To construct the tree, one begins at 
a site $i$ and selects the bond $b$ connected to $i$ with the lowest $e_{ij}$. 
Then, among all the {\it unvisited} bonds connected to $b$, the one with the 
lowest $e_{ij}$ is selected, and so on. This is not only the physical basis of
the RB model, but also for the {\it bond invasion percolation cluster} (BIPC) 
[10], if invasion is from a single site, with the role of $e_{ij}$ played by
the capillary pressure needed to enter a bond. The MST, or the BIPC, is a 
fractal object with [11,12], $D_f\simeq 1.22$ and 1.37 in 2D and 3D, 
respectively, compared with $D_f\simeq 1.89$ and 2.53 for the CPC. Thus, the 
question is, what is the appropriate $d_f$ in Eq. (1)?

Following SD's arguments [1], the correct $d_f$ is the harmonic dimension of 
the MSP or BIPC, not that of the CPC. Then, the relevant $d_H$ will not take 
on the same value of 4/3 in both 2D and 3D. Indeed, for the MST and BIPC $d_H$ 
takes on values very close to their $D_f$ mentioned above. Our extensive 
simulations confirm that ac conduction occurs in the MST or BIPC [13].

Note that when the SD's numerical results are fitted to Eq. (1), one obtains 
$d_f\simeq 1.35$, very close to 4/3, which might have motivated their 
conjecture. However, for 2D BIPC or the MST, $d_H\simeq 1.22$, smaller than 
4/3.

The structure of the MST or BIPC is universal, because only the {\it order} of 
the energies $e_{ij}$ matters, not their numerical values or their statistical 
distribution. Therefore, the universality of $\tilde{\sigma}$ is due to the 
universality of the BIPC or the MST. 

That ac conduction occurs in the MST, or the BIPC, also explains why the EMA 
provides accurate predictions for $\tilde{\sigma}$: the conduction cluster is a
low-dimensional, quasi-1D object, even in 3D. It is well-known that the EMA is 
very accurate for low-dimensional systems.

\bigskip

\bigskip

\newcounter{bean}
\begin{list}%
{[\arabic{bean}]}{\usecounter{bean}\setlength{\rightmargin}{\leftmargin}}

\item T.B. Schr{\o}der and J.C. Dyre, Phys. Rev. Lett. {\bf 101}, 025901
(2008).

\item J.C. Dyre and T.B. Schr{\o}der, Rev. Mod. Phys. {\bf 72}, 873 (2000).

\item M. Sahimi, M. Naderian, and F. Ebrahimi Phys. Rev. B {\bf 71}, 094208 
(2005). 

\item E. Pazhoohesh, H. Hamzehpour, and M. Sahimi, Phys. Rev. B {\bf 73}, 
174206 (2006).

\item V.V. Bryskin, Fix. Tverd. Tela (Leningrad) {\bf 22}, 2441 (1980)[Sov.
Phys. Solid State {\bf 22}, 1421 (1980)].

\item M. Sahimi, B.D. Hughes, L.E. Scriven, and H.T. Davis, J. Chem. Phys. 
{\bf 78}, 6849 (1983).

\item M. Sahimi, J. Phys. C {\bf 17}, 3957 (1984) explicitly derived $d_f=2$.

\item  J.C. Dyre, Phys. Rev. B {\bf 49}, 11709 (1994).

\item S. Alexander and R. Orbach, J. Phys. (Paris) Lett. {\bf 43}, L625 (1982).

\item D. Wilkinson and J. Willemsen, J. Phys. A {\bf 16}, 3365 (1983).

\item M. Sahimi, M. Hashemi, and J. Ghassemzadeh, Physica A {\bf 260}, 231 
(1999).

\item M. A. Knackstedt, M. Sahimi, and A.P. Sheppard, Phys. Rev. E {\bf 61}, 
4920 (2000).

\item M. Sahimi, to be published.

\end{list}%

\end{document}